\documentclass[prb,showpacs,superscriptaddress]{revtex4}
\usepackage{graphicx,amsmath,amssymb,natbib}

\begin{document}

\title{Osmotic compression of droplets of hard rods: \\A computer simulation study}

\author{Yu.\ Trukhina}
\affiliation{Institut f\"ur Physik, Johannes Gutenberg-Universit\"at,
D-55099 Mainz, Staudinger Weg 7, Germany}

\author{S.\ Jungblut}
\affiliation{Institut f\"ur Physik, Universit\"at Wien, Boltzmanngasse 5, 1090
  Wien, Austria}

\author{P.\ van\ der\ Schoot}
\affiliation{Faculteit Technische Natuurkunde, Technische Universiteit Eindhoven, Postbus 513, 5600 MB Eindhoven, The Netherlands}

\author{T.\ Schilling}
\affiliation{Institut f\"ur Physik, Johannes Gutenberg-Universit\"at,
D-55099 Mainz, Staudinger Weg 7, Germany}

\date{\today}

\begin{abstract}
By means of computer simulations we study how droplets of hard, rod-like
particles optimize their shape and internal structure under the influence of
the osmotic compression caused by the presence of spherical particles that act
as depletion agents. At sufficiently high osmotic pressures the rods that make
up the drops spontaneously align to turn them into uniaxial nematic liquid
crystalline droplets. The nematic droplets or ``tactoids'' that are formed
this way are not spherical but elongated, resulting from the competition between the anisotropic surface tension and the elastic deformation of the director field. In agreement with recent theoretical predictions we find that sufficiently small tactoids have a uniform director field, whilst large ones are characterized by a bipolar director field. From the shape and director-field transformation of the droplets we are able to estimate the surface anchoring strength and an average of the elastic constants of the hard-rod nematic.
\end{abstract}



\maketitle

\section{Introduction}

 Fluids of elongated particles that interact via a harshly repulsive excluded-volume interaction potential, i.e., hard-rod fluids, have proven immensely useful as a model system for investigating the formation, structure and properties of liquid crystals\cite{deGennes.Prost:1993}. Indeed, hard-rod fluids exhibit a wealth of aggregated states including nematic, smectic and columnar liquid-crystalline phases as well as various plastic and crystalline solid phases, depending on the precise shape, density and composition of the model particles present in the system\cite{bolhuis.frenkel:1997}.
Closest experimental realizations of the hard-rod fluid model are found in fluid dispersions of very stiff polymers, inorganic rod-like colloids,
filamentous viruses, fibrillar or tubular protein assemblies, and carbon
nanotubes\cite{Dogic:2006,Davidson:2005,Kas:1996,Zhang:2008,Vroege:1992}. For this class of system, the transition from an isotropic to a
symmetry-broken liquid-crystalline or crystalline state is entropy rather than
enthalpy driven, and the relevant control parameter is not the temperature but
the concentration or particle density. The transitions between the different
phases are driven by a tradeoff between free volume and configurational
entropy.

Arguably, the experimentally and theoretically by far most extensively studied liquid-crystalline state is the nematic that is characterized by uniaxial and up-down symmetry. The preferred mean
orientation of the particles is described by a director field the ground state
of which is spatially uniform. The transition from the isotropic to the
nematic phase is first order. (See, however, the work by Oakes and
co-workers \cite{Viamontes:2006}.) In the biphasic
region it proceeds via
the formation of nematic droplets called tactoids in the background
isotropic dispersion. Tactoids are typically not spherical but elongated and spindle-shaped, and they have over the past century or so been
observed in a wide variety of systems \cite{Sonin:1998, Bernal:1937,Bernal:1941,
  Zocher:1960, Dogic:2001,Oakes:2007,Kaznacheev:2003,Mourad:2008}. The shape and internal structure of nematic droplets is the
result of the competition between the preferred surface anchoring of the director field and the deformation of the director field that occurs if the preferred anchoring is indeed accommodated.

The shape and director-field configurations of tactoids have recently been studied theoretically in considerable detail within a macroscopic Frank
elasticity theory \cite{prinsen.schoot:2003, prinsen.schoot:2004a,
  prinsen.schoot:2004b,Kaznacheev:2002,Kaznacheev:2003}. Predictions depend on several parameters:
two surface energies (the surface tension and anchoring energy),
three bulk elastic constants associated with the splay,
twist and bend deformations, the saddle-splay surface elastic constant,
and, finally, the size of the droplet. For our purposes, it suffices to
summarize the most important predictions, presuming preferential planar anchoring of the director field to the interface that for rod-like particles seems to hold for entropy reasons: i) The director field of small drops is uniform
and of large ones bipolar; ii) The crossover is smooth
and set by a healing or extrapolation length that is the ratio of an average elastic constant and an anchoring energy; iii) The aspect ratio of
uniform tactoids depends only on
the anchoring energy, and of bipolar ones on their size relative to the healing length.

Some aspects of these theoretical predictions have been verified against
experiment, in particular the size dependence of aspect ratio and opening
angle of the sharp ends of the tactoids \cite{Kaznacheev:2002, Kaznacheev:2003,
  prinsen.schoot:2004a, prinsen.schoot:2004b, Oakes:2007}. The crossover from bipolar
to uniform director field has not been observed and this probably presents quite an experimental challenge because it has been predicted to
occur when the drops are quite small, that is, in the micrometer range.
For such small droplets, not at all very much larger than the particles that they are
made up from, director field patterns are difficult to distinguish in polarization microscopic images.

Clearly, computer simulations are of use here\cite{golemme:1988, erdmann:1990,
prishchepa:2005}, not least because a macroscopic
description might break down for such small drops, in other words, the predicted
transition may be spurious. On the other hand, recent simulations on the nucleation of the nematic phase in a hard-rod fluid have indeed shown that small nuclei of the nematic phase are elongated and
it appears at least that their director field is uniform \cite{Cuetos.Dijkstra:2007}. Simulations on small nematic droplets
of prolate particles interacting via a Gay-Berne potential have been shown to exhibit similar behavior, although fluctuations are large so a
director field is not so easily defined. \cite{Berardi:2003,Bates:2003} Because systematic computational
investigations of the shape and internal structure of nematic droplets
are lacking, we set out to do this for nematics of hard spherocylinders of aspect ratio eleven. Our simulations build on our earlier work on
hard-rod nematics in a spherical cavity of fixed size and shape \cite{trukhina:2008}, but we now focus on actual drops that can adjust their structure.
This we do by placing the spherocylinders in a suspension of spherical particles, which interact via a hard-core repulsion with the spherocylinders yet are transparent with
respect to themselves, i.e., behave like an ideal gas.

The method allows us to compress by osmotic stress
droplets of a fixed and predestined number of rod-like particles. This is
achieved in a controlled manner because the density of the
rods in the drop will be set by the imposed pressure of the ideal gas of spheres
(if we ignore a small contribution from a Laplace pressure). Hence, the drop size must then be given by
the number of rods in the simulation box. Note that our making use of what in essence is a depletion agent (the spherical particles)
is not all that dissimilar in spirit to the experiments of Dogic and Fraden\cite{Dogic:2001},
who used the non-adsorbing polymer dextran to induce  phase separation in aqueous dispersions of filamentous
fd virus particles.

Our results may be summarized as follows.
\begin{itemize}
\item[i)] For relatively low densities
of spheres, the droplets are spherical and the rods randomly oriented;
\item[ii)]  Above a critical sphere density, the drops become nematic and elongated;
\item[iii)] Small tactoids have a uniform director field and large ones
a bipolar one;
\item[iv)] At the crossover the drops measure about four rod lengths so indeed are very small;
\item[v)] The surface tension anisotropy and healing length that we
deduce from the simulations are consistent with theoretical estimates.
\end{itemize}

The remainder of this paper is organized as follows.
In section \ref{model} we introduce
the model and define the relevant order parameters.
A detailed discussion of our results is given in section \ref{res}. Finally, in section \ref{summary} we summarize our findings and relate
them to theory and experiment.

\section{Model and Simulation Details}
\label{model}

In the Monte-Carlo simulations, we model the rods as spherocylinders
consisting of a cylindrical part of length $L$ and diameter $D$, capped at both
ends by hemispheres of diameter $D$. In order to stabilize droplets of rods,
we surround them by a liquid of spheres.
The spherocylinders interact via a hard-core
repulsion. The spheres also interact with a hard-core repulsion with
spherocylinders but are able to freely interpenetrate
each other at not energy cost. Hence, our
model is an extension of the Asakura-Oosawa or AO model for colloid-polymer
mixtures \cite{asakura.oosawa:1954, vrij:1976} to anisotropic colloids \cite{Jungblut:2007}. Note that
strictly speaking the spheres
do interact with each other via the rods whilst this is not so in the original
AO model. 
This difference is of academic interest only, because we consider only
those concentrations of particles that produce a very
strong phase separation into sphere- and rod-rich phases, so
the former behaves as a (nearly) ideal gas
of spheres that in essence acts
as a barostat for the droplet consisting of only spherocylinders.

As already advertised in the introduction, this implies that by means of
changing the
number of spheres in the system at fixed volume,
we can tune the pressure exerted on the rods in the droplet. Through that we
vary the density of the drop and therefore also the physical properties
of the drop, such as the elastic response if the drop is in a nematic
fluid state. Hence, we use the number density of
spheres $\rho_{\rm{sp}}$ measured far away from the droplet as a
parameter
that characterizes the external conditions imposed on the droplets
formed. The same method has been used in an earlier study to study
the formation of nanocrystals in a simulation\cite{gruenwald:2007}.

The simulations were performed at fixed particle number $N$ and
simulation box volume $V$ (and temperature $T$, but as the system is purely
entropic, temperature is not relevant here) in a
cubic box with periodic boundary conditions.  We focus on
spherocylinders with an aspect ratio of $L/D+1=11$ (i.e. $L/D=10$), and spheres
with a diameter twice that of the thickness of the rods,
$D_{\rm sp}=2D$. Spheres that are much larger than this introduce strong
effects on the surface anchoring, whilst for smaller spheres the 
numbers needed in the simulation are so large to be unpractical from a computational point of view.

The number of spherocylinders in the box was varied from
$200$ to $700$, and the number of spheres was fixed such that
phase separation was induced into two phases containing virtually
only spherocylinders or spheres. This corresponded in
our simulations to between approximately 20000 and 70000 spheres.
We chose the simulation box to be $70^3 D^3$, i.e., sufficiently
large to ensure that the spherocylinders
did not interact directly with each other via the periodic boundaries.
We verified the droplets that form are
not system spanning. The systems were equilibrated by local translation and
rotation moves. Depending on the specific concentrations, $10^6 - 10^7$ MC 
sweeps were required.

The boundary and volume of a droplet is established as follows.
We first divide the system into small boxes and next verify whether a box contains the center of a sphere or whether it is intersected by any spherocylinder.
In the latter case, this box is counted as a part of the droplet.
If it so happens that a box does not fall into either of the two categories,
it is counted as a part of the droplet if it has a larger number of
nearest neighbouring boxes containing spherocylinders than spheres.

As is customary, we define the average alignment of the ($N$) rods in
terms of the traceless tensor $\bf{Q}$, with the elements
\[
\label{eq:s2}
 Q_{\alpha\beta} = \frac{1}{2 N} \sum_{i=1}^N
   \left( 3 u_{i\alpha} u_{i\beta} - \delta_{\alpha\beta} \right),
\]
where $u_{i\alpha}$ is the $\alpha$ component ($\alpha = x,y,z$)  of the
unit vector along the axis of particle $i=1,...,N$
 and $\delta_{\alpha\beta}$ the Kronecker delta.
Diagonalization of the tensor yields three eigenvalues, $\lambda_{+}$, $\lambda_{0}$ and
$\lambda_{-}$, where $\lambda_{+}>\lambda_{0}>\lambda_{-}$.
Different authors use different combinations of these eigenvalues to define
the nematic and biaxial order parameters\cite{dijkstra.roij.ea:2001, low:2002}. To avoid confusion,
we shall present all eigenvalues instead.
The case $\lambda_{+}>0$, $\lambda_{0}=\lambda_{-}$ corresponds to a structure with one preferred direction.
The case $\lambda_{+}=\lambda_{0}>0$ corresponds to a structure in which
one direction is avoided and the two other directions are equally favored. All the intermediate cases $\lambda_{+}>0$,
$\lambda_{+}>\lambda_{0}>\lambda_{-}$ correspond to a biaxial structure.
Obviously, in an isotropic phase one finds $\lambda_{+}=\lambda_{0}=\lambda_{-}=0$.

When calculating observables, such as the components of the 
order parameter tensor properly averaged over the ensemble of configurations,
all the possible types of symmetry in the system have to be taken into
account. In the case of radial symmetry we calculated the orientational
tensor directly by averaging over all configurations obtained in
the simulations. 

If there is a preferred axis in the system
along which the particles tend to align, we need to proceed differently. As 
rotations of the director do not cost any (free) energy,
the orientation of this axis can fluctuate strongly during a simulation run.
In order to average the local properties of interest, the configurations have been rotated in such a way that the director always points in the same direction. It is important to point out that this procedure
introduces additional noise to the simulation
results. In the following,
we show all data rotated such that the director is aligned with the $z$-axis.

\section{Results and Discussion}
\label{res}

\begin{figure}[ht]
\begin{center}
\includegraphics[height=5in, width=7in, angle=-0]
{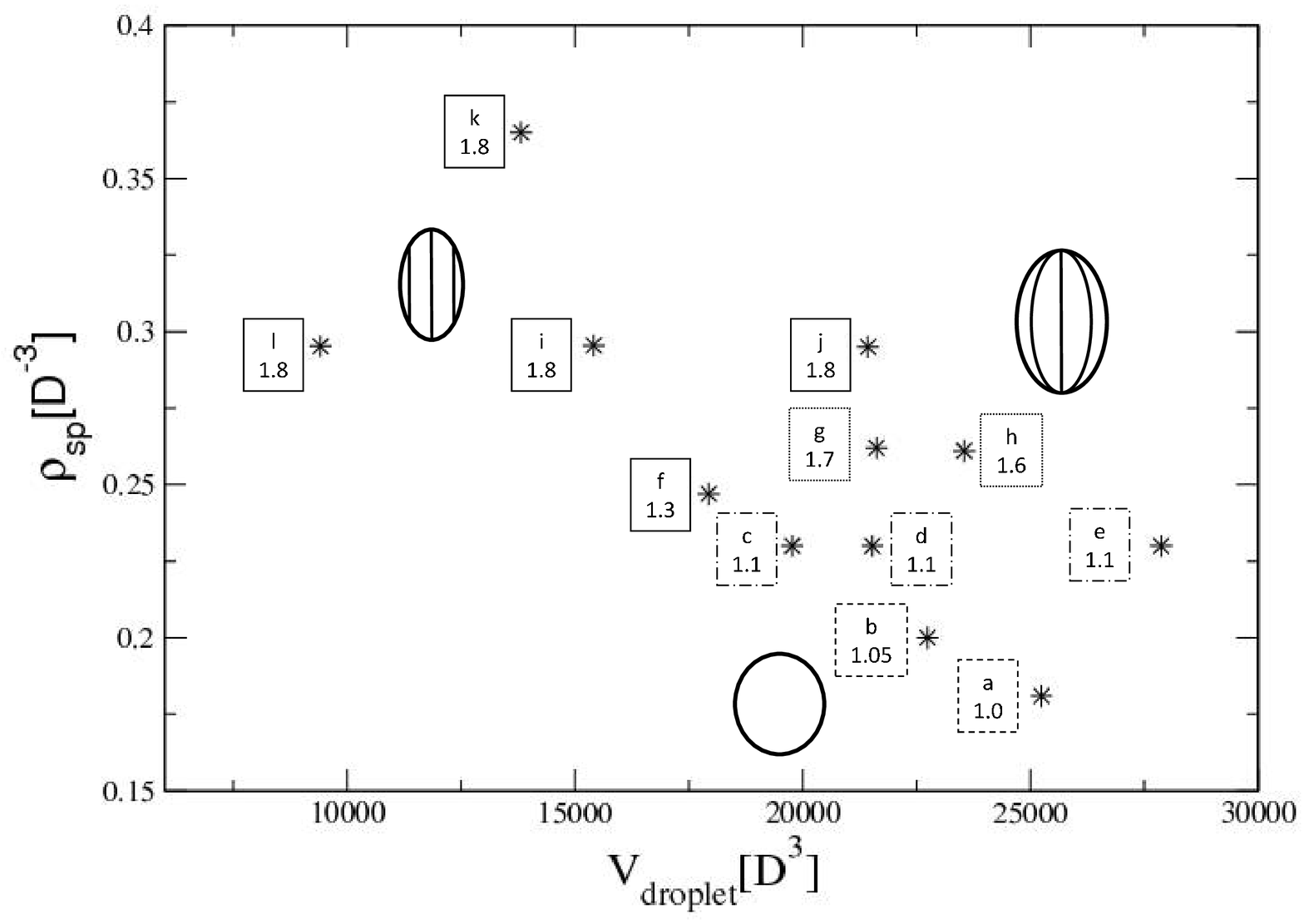} \caption{``Shape'' diagram of hard spherocylinder droplets immersed
in a fluid of spherical particles.
The horizontal axis shows the volume
of the droplet of spherocylinders in units of cylinder thickness cubed,
$D^3$, and the vertical axis
the number density of the spheres far away from the droplet.
The letters ``a'' to ``l'' are used in the main text to refer to specific points
on the diagram. The sketches show schematically the shape and the
interal structure of the droplets, and distinguish spherical isotropic
droplets, elongated nematic droplets with either a homogeneous or a
bipolar director field. The numbers in
the boxes indicate the aspect ratios of the droplets. The boundaries of the
boxes distinguish between: isotropic (dashed), transition region
(dash-dotted), bipolar (dotted) and homogeneous (solid).
} \label{diagram}
\end{center}
\end{figure}

In the diagram of fig.~\ref{diagram} we have indicated with asterisks
the system conditions for which we ran the simulations.
The letters ``a'' through ``l'' are used in the following to refer to specific
points in this diagram. We have indicated schematically
in the figure the shape and the structure of the droplets, which we
discuss in more detail below. The numbers in the
boxes indicate the aspect ratios of the droplets.

\begin{figure}[ht]
\begin{center}
\includegraphics[height=4in, width=2.8in, angle=-90]
{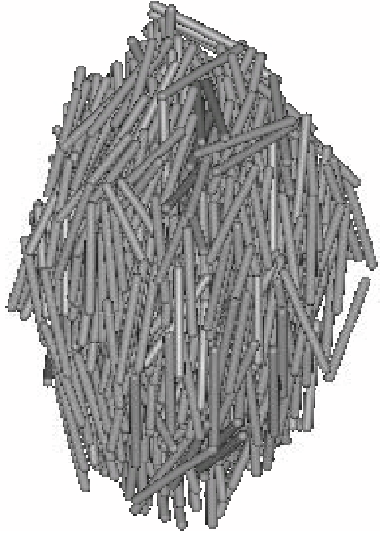} \caption{A snapshot of a nematic droplet of rods that forms for
 the conditions indicated by
 ``i'' in fig.~1. The spherical particles are not shown for clarity. The nematic total director is along the main axis of the drop. The
 average rod density in the drop is equal to $\rho D^3 = 0.034$.} \label{nem}
\end{center}
\end{figure}

\begin{figure}[ht]
\begin{center}
\includegraphics[height=4in, width=2.8in, angle=-90]
{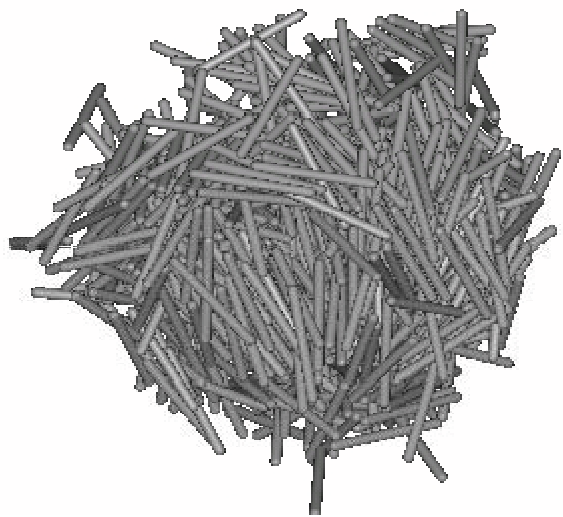} \caption{A snapshot of an almost isotropic droplet of rods for conditions ``c''. The average rod density in the drop is equal to $\rho D^3 = 0.027$.} \label{iso}
\end{center}
\end{figure}

Figs.~\ref{nem} and \ref{iso} give the snapshots of two typical structures given by conditions ``i'' and ``c'', respectively, where only the
spherocylinders are shown for clarity. Two essential
differences between these structures can be seen:
The first of the two droplets shown is i) more elongated
and ii) the spherocylinders in this droplet have a much stronger tendency to orient in one direction. Below we will analyse these effects in a more quantitative fashion.

The difference is caused by the difference in bulk sphere density and hence osmotic stress imposed on the drop by the sphere fluid.
Droplet ``i'' is subject to much higher
pressure and hence is condensed much more than droplet ``c'', and has
crossed over to the nematic phase.
The rod densities $\rho D^3$ in the two drops are $0.034$
and $0.027$ (averaged over each droplet).
From the (remarkably accurate) Lee-Parsons theory of the nematic transition
in bulk fluids of hard rods \cite{Cuetos:2007}, we expect for 
spherocylinders of aspect ratio 11 the nematic transition to occur at a 
pressure of $P D^3 /k_B T \approx 0.25 $, corresponding
to an ideal gas density of $\rho_{\rm sp}D^3 \approx 0.25$.
This is in reasonable agreement
with the transitional regime around $\rho_{\rm sp}D^3 \approx 0.23$ in the
diagram. We also estimated the transition density by
simulating a compression curve and an expansion curve in the bulk and obtained
 $\rho_{\rm sp}D^3 \approx 0.22$, which again is in good agreement with the
numbers above. We note that due to the effects of Laplace pressure the 
nematic transition should occur at a somewhat higher sphere density then
that in bulk solution. For details we refer to the Appendix. 

\begin{figure}[ht]
\begin{center}
\includegraphics[height=4in, width=4in, angle=-90]
{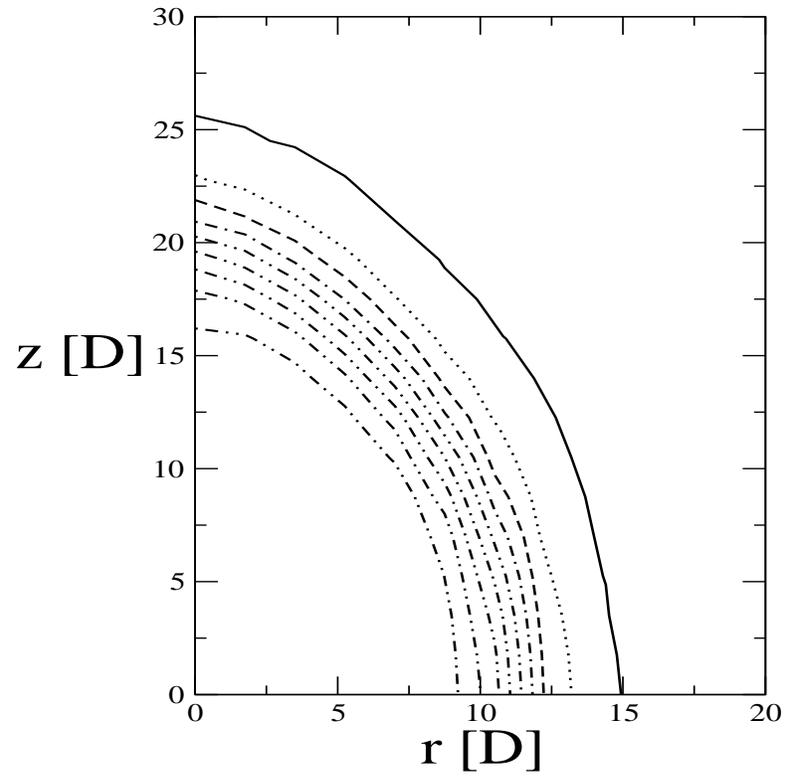} \caption{Isodensity lines of spherocylinders in a droplet from conditions ``i'', shown in cylindrical coordinates. The solid
line marks the boundary of the drop. Successive dashed, dash-dotted, etc., line
demarcate densities from $\rho D^3 = 0.039$ to $\rho D^3 = 0$.} \label{density}
\end{center}
\end{figure}

\begin{figure}[ht]
\begin{center}
\includegraphics[height=4in, width=4in, angle=-90]
{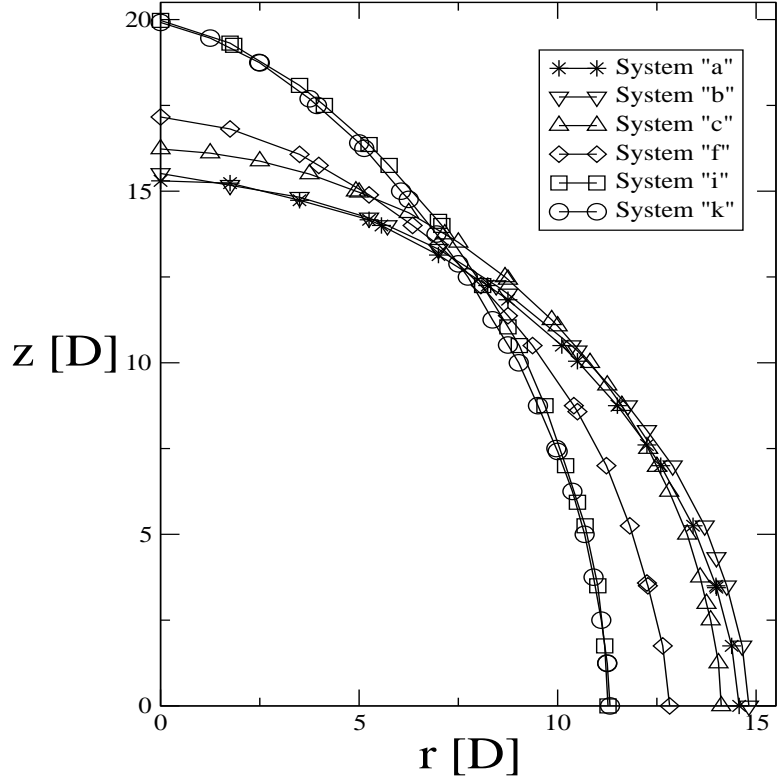} \caption{The shapes of droplets consisting of $500$ spherocylinders at different pressures of spheres (conditions ``a'', ``b'', ``c'', ``f'', ``i'', and ``k'' indicated in the schematic of fig.~ 1). Droplets ``a'' and ``b'' are those of isotropically oriented rods, and hence more or less spherical. Drop ``c'' is in the transition zone from the isotropic to the nematic state. Drops ``f'', ``i'' and ``k'' are nematic drops,
with ``i'' and ``k'' exhibiting more or less uniform director fields and
``f'' a more bipolar one.
.} \label{density_profiles}
\end{center}
\end{figure}

In fig.~\ref{density} we have plotted a typical iso-density distribution
in the $r$-$z$-plane for nematic droplet ``i''. Inside the droplet the spherocylinders have an approximately constant rod density equal to
 $\rho D^3 = 0.039$, which rapidly decays when approaching
 the interface to the fluid of spheres. It is clear that the drop is not spherical, the aspect ratio being approximately 1.8. For comparison,
 and in order to determine the shape of the droplets, we cut a slice from
the $r-z$-density profile at half of its maximum value. That is where we 
expect the Gibbs plane to be situated. The curves obtained in this way are shown in fig.~\ref{density_profiles} for several systems
consisting of the same amount or spherocylinders ($N_{\rm rods}=500$) but
with various densities of spheres, corresponding to the points
``a'',``b'',``c'',``f'',``i'', and ``k''.

At low pressures the droplets are (nearly) spherical. This is to be expected if the rods are in their isotropic state. However, if the
pressure is increased
the droplets crossover to the nematic phase and hence become elongated
in order to reduce
either the elastic deformation
of the director field if the anchoring is strong, or the anchoring free energy
if the anchoring is weak. As already advertised, the crossover occurs when the
typical drop dimension exceeds the healing length, which we are going to 
discuss in more depth in the following section.

On increase of the density of rods, the elastic constants in all likelihood
 increase, too, as should the interfacial tension and potentially also the anchoring strength.
We expect from scaling arguments that the ratio of the anchoring strength and the surface tension is a weak function of the pressure \cite{schoot:1999}, however, and in the weak anchoring regime it is this ratio that dictates the aspect ratio of the drops \cite{prinsen.schoot:2003}.  In the strong anchoring
regime the aspect ratio is an increasing function of the healing
length that in the equal-constant approximation
is given by the ratio of the elastic constant and the surface tension. Hence,
whether the droplets become more elongated then depends on how strongly these two energies depend on the pressure. Apparently, the elastic constants increase
more strongly with pressure because the aspect ratio increases from about 1.1 to 1.8
with increasing sphere concentration.

We observe the same tendency for the systems of 700 spherocylinders.
At low pressures the droplet is more or less spherical (state point ``e'', aspect ratio 1.1) and at high
pressures the droplet becomes elongated (state point ``j'', aspect ratio 1.8). We have to note, however,
that droplet size also affects the aspect ratio: it decreases with increasing size.
This has been observed experimentally and is predicted theoretically based
on macroscopic theory\cite{Kaznacheev:2003,
  prinsen.schoot:2004b, Oakes:2007}.
The aspect ratios for all the state points are shown in the boxes in the ``shape'' diagram fig.~\ref{diagram}, with an estimated error of $0.1$.

\begin{figure}[ht]
\begin{center}
\includegraphics[height=6in, width=4.2in, angle=-90]
{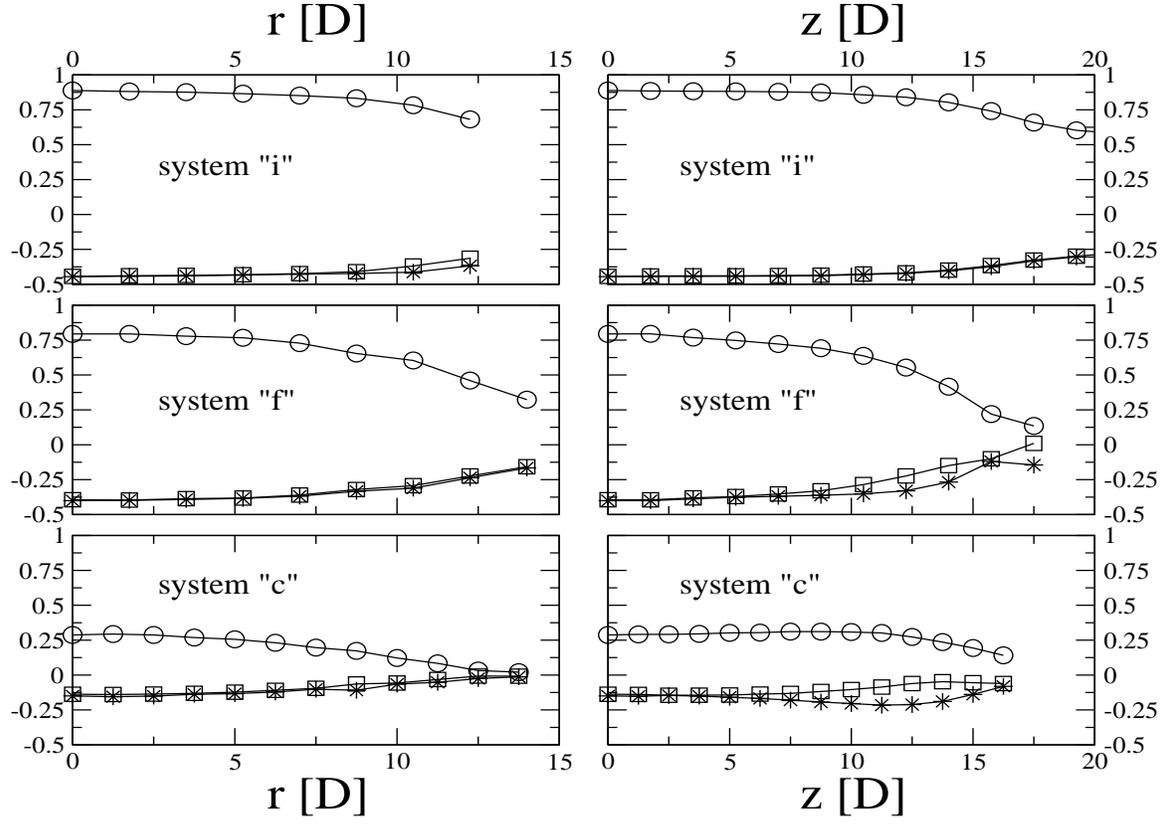} \caption{The r- and z-profiles of the eigenvalues of the orientational tensor $Q$ for the systems "i", "f", and "c" (each consisting of 500 spherocylinders).} \label{eigenvalues}
\end{center}
\end{figure}

We now turn to the orientational state of the
rods in the droplets.
This is described in
terms of the eigenvalues of the orientational tensor $\bf{Q}$
(see \ref{model}). In fig.~\ref{eigenvalues} we show the r- and z-profiles
of these eigenvalues for the drops ``i'', ``f'', and ``c'' (that each consist
of 500 spherocylinders). The droplets become less ordered
and less elongated.
Droplet ``i'' has a high nematic ordering in its center, which slightly
decays on approach of the interface to the gas of spheres in both, 
the r- and the z-direction.
Droplet ``f'' has a lower density and also
a nematic order parameter that is smaller.
Droplet ``c'' has even lower values of the nematic order parameter, and
at the interfaces all the eigenvalues become equal to $0$, indicating
an isotropic configuration. This drop probably is close to the conditions where
the isotropic-nematic transition takes place.

By analysing the tensor $\bf{Q}$ for all the investigated droplets, we construct scalar order parameter profiles as well as
the nematic director-field configuration. The results are indicated
schematically in the
diagram \ref{diagram}:
\begin{itemize}
\item[1)] Droplets ``a'' and ``b'' are spherical
 droplets of an isotropic rod fluid;
\item[2)] Droplets ``c'', ``d'' and ``e'' are in
the isotropic-nematic transition region, exhibiting a strongly
fluctuating orientational order;
\item[3)] States ``f'' to ``l'' are strongly nematic drops;
\item[4)] The director field of drops ``i'', ``j'', ``k'' and ``l'' is more or less uniform,
those of ``g'' and``h'' bipolar, and of ``f'' in between these two.
\end{itemize}

\begin{figure}[ht]
\begin{center}
\includegraphics[height=4in, width=2.8in, angle=-0]
{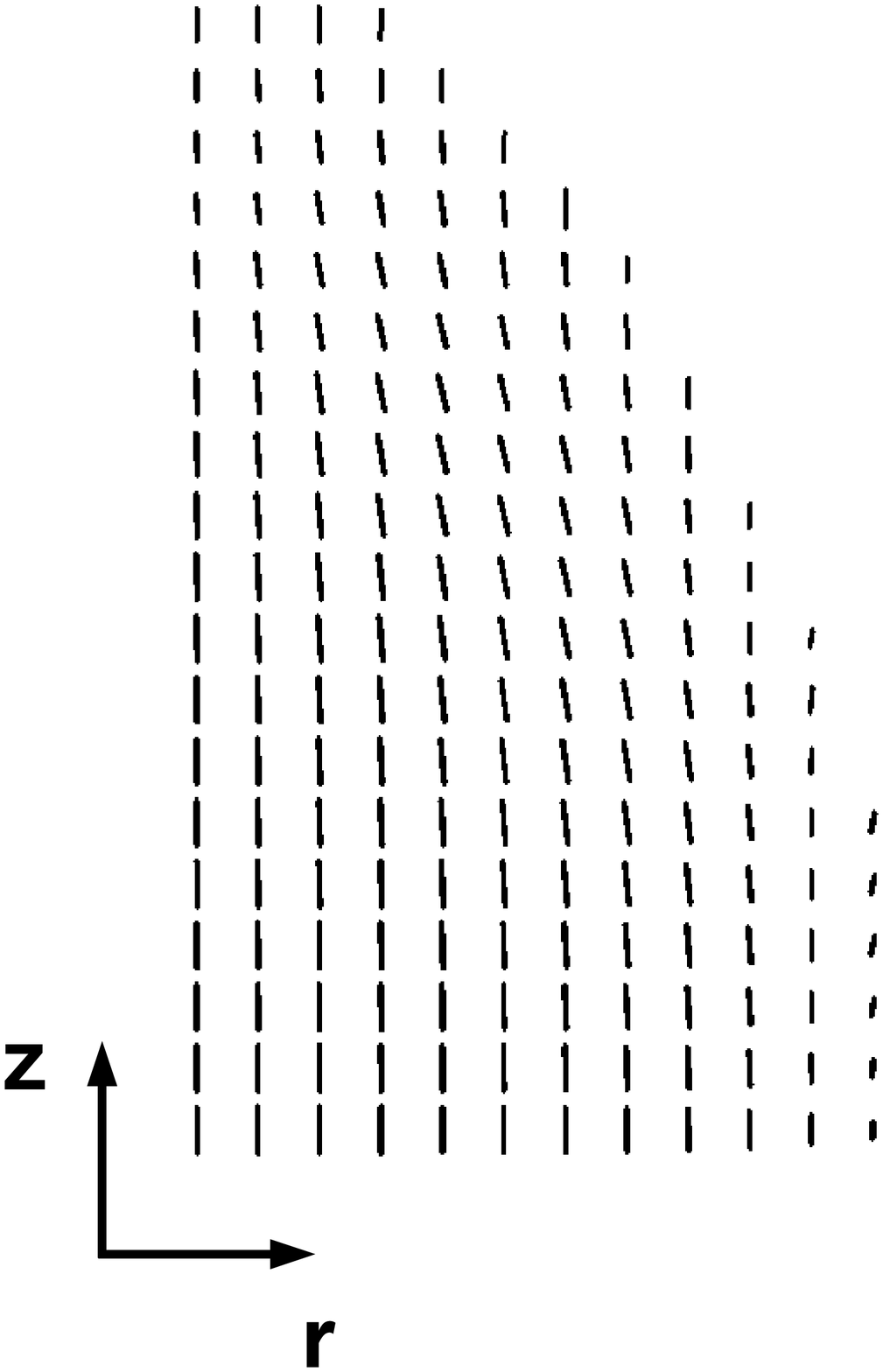} \caption{An example of a nematic droplet with a homogeneous
  director field (system "k"), the eigenvectors corresponging to the maximum
  eigenvalues of the orientational tensor ${\bf Q}$ are shown in polar coordinates.} \label{S2_homog}
\end{center}
\end{figure}

\begin{figure}[ht]
\begin{center}
\includegraphics[height=4in, width=2.8in, angle=-0]
{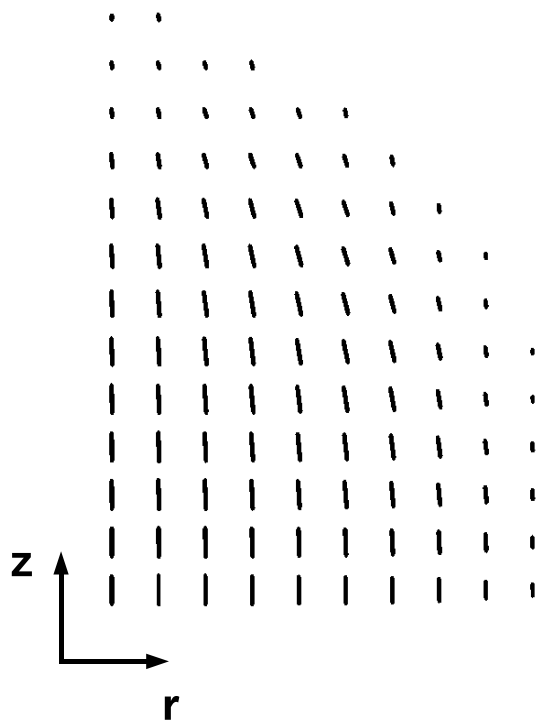} \caption{An example of a nematic droplet with a bipolar director field (system "h"), the eigenvectors corresponging to the maximum eigenvalues of the orientational tensor $Q$ are shown in polar coordinates.} \label{S2_bip}
\end{center}
\end{figure}

To illustrate these findings, we show the director field (the axis
given by the eigenvector corresponding to the largest eigenvalue of
the orientation tensor)
as a function
of the radial and axial distances $\bf{r}$ and $\bf{z}$ in fig.~\ref{S2_homog} for the case ``k'' and in fig.~\ref{S2_bip} for the case ``h''. The director
field of the tactoid ``k'' is more or less uniform and oriented
along the main axis of the drop. That of ``h'' is bipolar, i.e., curved along the elongated drop surface toward
the tips, where the scalar order parameter drops to zero. This signifies
the melting of the nematic near the tips, where theoretically one would expect
the surface point defects (``boojums'') to reside \cite{prinsen.schoot:2003}.

The director field in a nematic droplet is determined by the interplay of
surface anchoring and elastic forces. The bipolar structure can only be formed if the energy for bending is small enough compared to
the surface energy. This can be achieved
if the density of the suspension is small (but the droplet still
has a nematic
structure), or if the droplet is big (and, therefore, the curvature of the
interface is small). Systems ``g'' and ``h'' fit into this category.

The aspect ratio of the droplets does not depend on their size if the director
field inside the droplets is homogeneous. This follows directly
from the well-known Wulff construction of the droplet shape given any polar
angle-dependent
surface tension \cite{prinsen.schoot:2003}. Indeed, all the droplets with an approximately
homogeneous director field, droplets ``i'', ``j'', ``k'', and ``l'',
that we obtained
in our simulations have the same aspect ratio of 1.8
within a statistical error of the simulation of about 0.1,
in agreement with this theoretical prediction.
From the aspect ratio observed in the simulations we can in fact deduce 
a dimensionless anchoring strength. Let us presume
that the anisotropic surface
tension $\gamma$  has a functional form
of the Rapini-Papoular type, so
$\gamma = \tau [1+\omega (\bf{q} \cdot \bf{n})^2]$, with $\tau$ the bare surface
tension, $\omega$ the dimensionless anchoring strength, $\bf{q}$ the
surface normal and $\bf{n}$ the director field at the
surface of the drop. For planar anchoring to be favored, $\omega > 0$.
From the Wulff construction we then deduce that the aspect ratio of
the drop equals $1+\omega$ for $0\leq \omega \leq 1$ and $2\sqrt{\omega}$ for $\omega>1$ \cite{prinsen.schoot:2003}.
Hence, we find from our simulations a value for $\omega$ of 0.8, quite close the value of 0.65 found by Dijkstra and co-workers in a simulation
study of nematic drops nucleated in a super-saturated dispersion of hard
rods\cite{cuetos:2008}.

Finally, from the crossover from bipolar to homogeneous director fields, we obtain an estimate for the healing length. According to fig.~ 1,
the crossover occurs for drops of a volume about  $1.7\times 10^4 D^3$. This value may actually depend on the
sphere density, but because of the lack of any detailed information
we shall ignore this for
simplicity. Presuming that the bend elastic constant is
about ten times larger than the splay elastic constant, which seems 
reasonable on account of predictions for hard rods in the Onsager limit, 
the crossover occurs at a droplet
volume equal to about 10 times the healing length $\lambda \equiv
(K_{11}-K_{24})/\tau \omega$ cubed according
to macroscopic theory \cite{prinsen.schoot:2004a, prinsen.schoot:2004b}, 
where $K_{11}$ denotes the splay elastic constant and $K_{24}$
the saddle-splay surface elastic constant. So, we find for the healing length $\lambda \approx 12 D$, which is about a rod length.
Clearly, macroscopic theories, such as those of Kaznacheev and collaborators
\cite{Kaznacheev:2003, Kaznacheev:2002}, and of Prinsen and
van der Schoot \cite {prinsen.schoot:2003, prinsen.schoot:2004a,
  prinsen.schoot:2004b}, could perhaps be expected to break down at such 
small length scales, yet the predicted crossover from uniform to bipolar 
director fields apparently still survives.

From the estimate of the healing length, we can obtain an order of
magnitude estimate of the interfacial tension between the rods
and the spheres that we can compare with the scaling estimate
given in the Appendix. 
If we presume the Saupe-Nehring relation to hold, implying that
$K_{24} = (K_{11}-K_{22})/2$ $\;$\cite{Nehring:1971}, and make use of the approximate
expression $K_{22}=K_{11}/3$ $\;$\cite{Lee:1986}, we obtain $\lambda \approx 2K_{11}/3\tau \omega$. Hence, $K_{11}/\tau \approx 14 D$, or, $\beta \tau D^2 \approx \beta K_{11} D/14$. Within
a second-virial approximation, which admittedly is not very accurate
for rods of aspect ratio below 20, we expect $\beta K_{11} D \approx 0.9$ to hold near the transition \cite{Vroege:1987}. Hence, for the interfacial tension between the rods and the
spheres we obtain the estimate $\beta \tau D^2 \approx 0.07$. According to our
scaling estimate cited in the Appendix, we have $\beta \tau D^2 \approx 0.25
\alpha \xi/D$ with $\alpha$ a prefactor that should be of order 0.1
\cite{schoot:1999} and $\xi$ the interfacial width. If $\xi \approx L$, this then implies that $\alpha \approx 0.03$, which is somewhat smaller than expected \cite{schoot:1999}.

In order to go beyond this quantitative analysis, simulation data of the 
elastic constants of the bulk nematic and of the surface tension between 
the co-existing bulk fluids would be necessary. Unfortunately, these are not
available yet for spherocylinders of aspect ratio 11, as, in particular, 
simulations to determine elastic constants are computationally rather 
expensive. (To our knowledge, elastic constants have been computed only for 
spherocylinders of aspect ratio 6 \cite{Tjipto:1992}.)

\section{Summary}
\label{summary}
By means of computer simulations we have shown that fluid droplets of hard
rods, osmotically compressed by the presence of spherical particles, undergo
an isotropic-nematic transition at sufficiently high osmotic stress. We find
the nematic droplets not to be spherical but elongated. The director field of
the drops is uniform if smaller than a critical size and bipolar if larger
than that. We interpret our findings in terms of the predictions of continuum
mechanical theory that minimizes the combined effect of an elastic deformation
of the director field and an anchoring frustration of this director field at
the surface of the drops. Although in our simulations the drops are not at all
large on the scale of the rods, and continuum theory should perhaps not be
expected to be accurate, results from both levels of description seem to be
consistent with each other down to drop sizes that are as small as a few times the particle length.

\section{Appendix}
\label{Appendix}
An estimate for the rod density in the droplet as a function of the number
$N_{\rm rods}$ of rods and the density of spheres $\rho_{\rm sp}$ can be obtained by presuming complete demixing of the two components and by presuming that the interface between them is sharp. Let $R$ be the radius of the drop, assumed perfectly
spherical, then
$\rho =3 N_{\rm rods}/4\pi R^3 $ is the density of the rods in drop. The bulk pressure of the hard rods is to a very good approximation equal to the expression
put forward by Parsons and by Lee \cite{Vroege:1992}
\[
\beta P = \rho \left(1 + \frac{2\phi(2-\phi)}{(1-\phi)^3}\left[
1+\frac{3\pi}{8}\Lambda\right]\right),
\]
at least in the isotropic phase, where $\phi = \rho (\pi D^3/6 + \pi LD^2/4)$ denotes the packing fraction, $\Lambda = (L/D)^2/\pi(1+3L/2D)$ for slender rods is
proportional to their aspect ratio and $\beta$ denotes the reciprocal thermal
energy $1/k_BT$ with $k_B$ Boltzmann's constant and $T$ the absolute
temperature. The pressure of the ideal gas of spheres obeys
\[
\beta P_{\rm sp} = \rho_{\rm sp}.
\]
Mechanical equilibrium between the gas of hard rods and that of ideal spheres demands that
\[
P + \frac{2\gamma}{R}= P_{\rm sp},
\]
where the second term on the left-hand side is the contribution from the Laplace pressure across the curved interface, with $\gamma$ the interfacial tension that presumably depends on the bulk densities of both the rods and the spheres.
For any given number of rods $N_{\rm rods}$, this equation sets the equilibrium size of the drop.

An estimate of the magnitude of the Laplace pressure may be given by making
use of the scaling Ansatz $\gamma \approx P_{\rm sp}\xi$, with
$\xi \approx L$ the actual interfacial width \cite{schoot:1999}. Hence,
$P/P_{\rm sp} \approx 1 - 2\alpha\xi /R$ with $\alpha$ a constant of proportionality
that we estimate to be of order 0.1 $\;$ \cite{schoot:1999}. So,
the presence of the interface reduces the pressure
in the drop relative to that in the reservoir of spheres
and hence postpones the onset of the nematic phase to higher densities
of spheres the smaller the drop.

From Table I of Lee \cite{Lee:1987}, we deduce by linear extrapolation that for rods of $L/D=10$ the
bulk nematic phase sets in at a dimensionless pressure
$\beta P D^3 \approx 0.247$, corresponding to a sphere fraction of $\rho_{\rm sp} D^3 \approx
0.247$, in reasonable agreement with what we find in the simulations. See fig.~ 1.

\acknowledgments We thank K.~Binder and M.~Allen
for helpful suggestions. CPU time was provided on the JUMP by the John
von Neumann Centre in J\"ulich. We thank the Deutsche Forschungsgemeinschaft
(DFG, Emmy Noether Program and SFB Tr6) and the MWFZ Mainz for financial support.

\newcommand{\noopsort}[1]{} \newcommand{\printfirst}[2]{#1}
  \newcommand{\singleletter}[1]{#1} \newcommand{\switchargs}[2]{#2#1}

\clearpage

\clearpage

\clearpage

\clearpage

\clearpage

\clearpage

\clearpage

\clearpage


\begin{thebibliography}{41}
\expandafter\ifx\csname natexlab\endcsname\relax\def\natexlab#1{#1}\fi
\expandafter\ifx\csname bibnamefont\endcsname\relax
  \def\bibnamefont#1{#1}\fi
\expandafter\ifx\csname bibfnamefont\endcsname\relax
  \def\bibfnamefont#1{#1}\fi
\expandafter\ifx\csname citenamefont\endcsname\relax
  \def\citenamefont#1{#1}\fi
\expandafter\ifx\csname url\endcsname\relax
  \def\url#1{\texttt{#1}}\fi
\expandafter\ifx\csname urlprefix\endcsname\relax\def\urlprefix{URL }\fi
\providecommand{\bibinfo}[2]{#2}
\providecommand{\eprint}[2][]{\url{#2}}

\bibitem[{\citenamefont{de~Gennes and Prost}(1993)}]{deGennes.Prost:1993}
\bibinfo{author}{\bibfnamefont{P.~G.} \bibnamefont{de~Gennes}}
  \bibnamefont{and} \bibinfo{author}{\bibfnamefont{J.}~\bibnamefont{Prost}},
  \emph{\bibinfo{title}{The Physics of Liquid Crystals}}
  (\bibinfo{publisher}{Clarendon Press}, \bibinfo{address}{Oxford},
  \bibinfo{year}{1993}).

\bibitem[{\citenamefont{Bolhuis and Frenkel}(1997)}]{bolhuis.frenkel:1997}
\bibinfo{author}{\bibfnamefont{P.}~\bibnamefont{Bolhuis}} \bibnamefont{and}
  \bibinfo{author}{\bibfnamefont{D.}~\bibnamefont{Frenkel}},
  \bibinfo{journal}{J. Chem. Phys.} \textbf{\bibinfo{volume}{106}},
  \bibinfo{pages}{666} (\bibinfo{year}{1997}).

\bibitem[{\citenamefont{Dogic and Fraden}(2006)}]{Dogic:2006}
\bibinfo{author}{\bibfnamefont{Z.}~\bibnamefont{Dogic}} \bibnamefont{and}
  \bibinfo{author}{\bibfnamefont{S.}~\bibnamefont{Fraden}},
  \bibinfo{journal}{Current Opinion in Colloid {\&} Interface Science}
  \textbf{\bibinfo{volume}{11}}, \bibinfo{pages}{47} (\bibinfo{year}{2006}).

\bibitem[{\citenamefont{Davidson and Gabriel}(2005)}]{Davidson:2005}
\bibinfo{author}{\bibfnamefont{P.}~\bibnamefont{Davidson}} \bibnamefont{and}
  \bibinfo{author}{\bibfnamefont{J.~C.~P.} \bibnamefont{Gabriel}},
  \bibinfo{journal}{Current Opinion in Colloid {\&} Interface Science}
  \textbf{\bibinfo{volume}{9}}, \bibinfo{pages}{377} (\bibinfo{year}{2005}).

\bibitem[{\citenamefont{Kas et~al.}(1996)\citenamefont{Kas, Strey, Tang,
  Finger, Ezzell, and Sackmann}}]{Kas:1996}
\bibinfo{author}{\bibfnamefont{J.}~\bibnamefont{Kas}},
  \bibinfo{author}{\bibfnamefont{H.}~\bibnamefont{Strey}},
  \bibinfo{author}{\bibfnamefont{J.~X.} \bibnamefont{Tang}},
  \bibinfo{author}{\bibfnamefont{D.}~\bibnamefont{Finger}},
  \bibinfo{author}{\bibfnamefont{R.}~\bibnamefont{Ezzell}}, \bibnamefont{and}
  \bibinfo{author}{\bibfnamefont{E.}~\bibnamefont{Sackmann}},
  \bibinfo{journal}{Biophys. J.} \textbf{\bibinfo{volume}{70}},
  \bibinfo{pages}{609} (\bibinfo{year}{1996}).

\bibitem[{\citenamefont{Zhang and Kumar}(2008)}]{Zhang:2008}
\bibinfo{author}{\bibfnamefont{S.~J.} \bibnamefont{Zhang}} \bibnamefont{and}
  \bibinfo{author}{\bibfnamefont{S.}~\bibnamefont{Kumar}},
  \bibinfo{journal}{Small} \textbf{\bibinfo{volume}{4}}, \bibinfo{pages}{1270}
  (\bibinfo{year}{2008}).

\bibitem[{\citenamefont{Vroege and Lekkerkerker}(1992)}]{Vroege:1992}
\bibinfo{author}{\bibfnamefont{G.~J.} \bibnamefont{Vroege}} \bibnamefont{and}
  \bibinfo{author}{\bibfnamefont{H.}~\bibnamefont{Lekkerkerker}},
  \bibinfo{journal}{Rep. Progr. Phys.} \textbf{\bibinfo{volume}{55}},
  \bibinfo{pages}{1241} (\bibinfo{year}{1992}).

\bibitem[{\citenamefont{Viamontes et~al.}(2006)\citenamefont{Viamontes, Oakes,
  and Tang}}]{Viamontes:2006}
\bibinfo{author}{\bibfnamefont{J.}~\bibnamefont{Viamontes}},
  \bibinfo{author}{\bibfnamefont{P.~W.} \bibnamefont{Oakes}}, \bibnamefont{and}
  \bibinfo{author}{\bibfnamefont{J.~X.} \bibnamefont{Tang}},
  \bibinfo{journal}{Phys. Rev. Lett.} \textbf{\bibinfo{volume}{97}},
  \bibinfo{pages}{118103} (\bibinfo{year}{2006}).

\bibitem[{\citenamefont{Sonin}(1998)}]{Sonin:1998}
\bibinfo{author}{\bibfnamefont{A.~S.} \bibnamefont{Sonin}},
  \bibinfo{journal}{Colloid J. USSR} \textbf{\bibinfo{volume}{60}},
  \bibinfo{pages}{129} (\bibinfo{year}{1998}).

\bibitem[{\citenamefont{Bernal and Fankuchen}(1937)}]{Bernal:1937}
\bibinfo{author}{\bibfnamefont{J.~D.} \bibnamefont{Bernal}} \bibnamefont{and}
  \bibinfo{author}{\bibfnamefont{I.}~\bibnamefont{Fankuchen}},
  \bibinfo{journal}{Nature (London)} \textbf{\bibinfo{volume}{139}},
  \bibinfo{pages}{923} (\bibinfo{year}{1937}).

\bibitem[{\citenamefont{Bernal and Frankuchen}(1941)}]{Bernal:1941}
\bibinfo{author}{\bibfnamefont{J.~D.} \bibnamefont{Bernal}} \bibnamefont{and}
  \bibinfo{author}{\bibfnamefont{I.}~\bibnamefont{Frankuchen}},
  \bibinfo{journal}{J. Gen. Physiol.} \textbf{\bibinfo{volume}{25}},
  \bibinfo{pages}{111} (\bibinfo{year}{1941}).

\bibitem[{\citenamefont{Zocher and Toeroek}(1960)}]{Zocher:1960}
\bibinfo{author}{\bibfnamefont{H.}~\bibnamefont{Zocher}} \bibnamefont{and}
  \bibinfo{author}{\bibfnamefont{C.}~\bibnamefont{Toeroek}},
  \bibinfo{journal}{Kolloid-Z.} \textbf{\bibinfo{volume}{170}},
  \bibinfo{pages}{140} (\bibinfo{year}{1960}).

\bibitem[{\citenamefont{Dogic and Fraden}(2001)}]{Dogic:2001}
\bibinfo{author}{\bibfnamefont{Z.}~\bibnamefont{Dogic}} \bibnamefont{and}
  \bibinfo{author}{\bibfnamefont{S.}~\bibnamefont{Fraden}},
  \bibinfo{journal}{Philos. Trans. R. Soc. London, Ser. A}
  \textbf{\bibinfo{volume}{359}}, \bibinfo{pages}{997} (\bibinfo{year}{2001}).

\bibitem[{\citenamefont{Oakes et~al.}(2007)\citenamefont{Oakes, Viamontes, and
  Tang}}]{Oakes:2007}
\bibinfo{author}{\bibfnamefont{P.~W.} \bibnamefont{Oakes}},
  \bibinfo{author}{\bibfnamefont{J.}~\bibnamefont{Viamontes}},
  \bibnamefont{and} \bibinfo{author}{\bibfnamefont{J.~X.} \bibnamefont{Tang}},
  \bibinfo{journal}{Phys. Rev. E} \textbf{\bibinfo{volume}{75}},
  \bibinfo{pages}{061902} (\bibinfo{year}{2007}).

\bibitem[{\citenamefont{Kaznacheev et~al.}(2003)\citenamefont{Kaznacheev,
  Bogdanov, and Sonin}}]{Kaznacheev:2003}
\bibinfo{author}{\bibfnamefont{A.~V.} \bibnamefont{Kaznacheev}},
  \bibinfo{author}{\bibfnamefont{M.~M.} \bibnamefont{Bogdanov}},
  \bibnamefont{and} \bibinfo{author}{\bibfnamefont{A.~S.} \bibnamefont{Sonin}},
  \bibinfo{journal}{J. Exp. Theor. Phys.} \textbf{\bibinfo{volume}{97}},
  \bibinfo{pages}{1159} (\bibinfo{year}{2003}).

\bibitem[{\citenamefont{Mourad et~al.}(2008)\citenamefont{Mourad, Devid, van
  Schooneveld, Vonk, and Lekkerkerker}}]{Mourad:2008}
\bibinfo{author}{\bibfnamefont{M.~C.~D.} \bibnamefont{Mourad}},
  \bibinfo{author}{\bibfnamefont{E.~J.} \bibnamefont{Devid}},
  \bibinfo{author}{\bibfnamefont{M.~M.} \bibnamefont{van Schooneveld}},
  \bibinfo{author}{\bibfnamefont{C.}~\bibnamefont{Vonk}}, \bibnamefont{and}
  \bibinfo{author}{\bibfnamefont{H.~N.~W.} \bibnamefont{Lekkerkerker}},
  \bibinfo{journal}{J. Chem. Phys. B} \textbf{\bibinfo{volume}{112}},
  \bibinfo{pages}{10142} (\bibinfo{year}{2008}).

\bibitem[{\citenamefont{Prinsen and {van der
  Schoot}}(2003)}]{prinsen.schoot:2003}
\bibinfo{author}{\bibfnamefont{P.}~\bibnamefont{Prinsen}} \bibnamefont{and}
  \bibinfo{author}{\bibfnamefont{P.}~\bibnamefont{{van der Schoot}}},
  \bibinfo{journal}{Phys. Rev. E} \textbf{\bibinfo{volume}{68}},
  \bibinfo{pages}{021701} (\bibinfo{year}{2003}).

\bibitem[{\citenamefont{Prinsen and {van der
  Schoot}}(2004{\natexlab{a}})}]{prinsen.schoot:2004a}
\bibinfo{author}{\bibfnamefont{P.}~\bibnamefont{Prinsen}} \bibnamefont{and}
  \bibinfo{author}{\bibfnamefont{P.}~\bibnamefont{{van der Schoot}}},
  \bibinfo{journal}{Eur. Phys. J. E} \textbf{\bibinfo{volume}{13}},
  \bibinfo{pages}{35} (\bibinfo{year}{2004}{\natexlab{a}}).

\bibitem[{\citenamefont{Prinsen and {van der
  Schoot}}(2004{\natexlab{b}})}]{prinsen.schoot:2004b}
\bibinfo{author}{\bibfnamefont{P.}~\bibnamefont{Prinsen}} \bibnamefont{and}
  \bibinfo{author}{\bibfnamefont{P.}~\bibnamefont{{van der Schoot}}},
  \bibinfo{journal}{J. Phys.: Condens. Matter} \textbf{\bibinfo{volume}{16}},
  \bibinfo{pages}{8835} (\bibinfo{year}{2004}{\natexlab{b}}).

\bibitem[{\citenamefont{Kaznacheev et~al.}(2002)\citenamefont{Kaznacheev,
  Bogdanov, and Taraskin}}]{Kaznacheev:2002}
\bibinfo{author}{\bibfnamefont{A.~V.} \bibnamefont{Kaznacheev}},
  \bibinfo{author}{\bibfnamefont{M.~M.} \bibnamefont{Bogdanov}},
  \bibnamefont{and} \bibinfo{author}{\bibfnamefont{S.~A.}
  \bibnamefont{Taraskin}}, \bibinfo{journal}{J. Exp. Theor. Phys.}
  \textbf{\bibinfo{volume}{95}}, \bibinfo{pages}{57} (\bibinfo{year}{2002}).

\bibitem[{\citenamefont{Golemme et~al.}(1988)\citenamefont{Golemme, \u{Z}umer,
  Allender, and Doane}}]{golemme:1988}
\bibinfo{author}{\bibfnamefont{A.}~\bibnamefont{Golemme}},
  \bibinfo{author}{\bibfnamefont{S.}~\bibnamefont{\u{Z}umer}},
  \bibinfo{author}{\bibfnamefont{D.~W.} \bibnamefont{Allender}},
  \bibnamefont{and} \bibinfo{author}{\bibfnamefont{J.~W.} \bibnamefont{Doane}},
  \bibinfo{journal}{Phys. Rev. Lett.} \textbf{\bibinfo{volume}{61}},
  \bibinfo{pages}{2937} (\bibinfo{year}{1988}).

\bibitem[{\citenamefont{Erdmann et~al.}(1990)\citenamefont{Erdmann, \u{Z}umer,
  and Doane}}]{erdmann:1990}
\bibinfo{author}{\bibfnamefont{J.~H.} \bibnamefont{Erdmann}},
  \bibinfo{author}{\bibfnamefont{S.}~\bibnamefont{\u{Z}umer}},
  \bibnamefont{and} \bibinfo{author}{\bibfnamefont{J.~W.} \bibnamefont{Doane}},
  \bibinfo{journal}{Phys. Rev. Lett.} \textbf{\bibinfo{volume}{64}},
  \bibinfo{pages}{1907} (\bibinfo{year}{1990}).

\bibitem[{\citenamefont{Prishchepa et~al.}(2005)\citenamefont{Prishchepa,
  Shabanov, and Zyryanov}}]{prishchepa:2005}
\bibinfo{author}{\bibfnamefont{O.~O.} \bibnamefont{Prishchepa}},
  \bibinfo{author}{\bibfnamefont{A.~V.} \bibnamefont{Shabanov}},
  \bibnamefont{and} \bibinfo{author}{\bibfnamefont{V.~Y.}
  \bibnamefont{Zyryanov}}, \bibinfo{journal}{Phys. Rev. E}
  \textbf{\bibinfo{volume}{72}}, \bibinfo{pages}{031712}
  (\bibinfo{year}{2005}).

\bibitem[{\citenamefont{Cuetos and Dijkstra}(2007)}]{Cuetos.Dijkstra:2007}
\bibinfo{author}{\bibfnamefont{A.}~\bibnamefont{Cuetos}} \bibnamefont{and}
  \bibinfo{author}{\bibfnamefont{M.}~\bibnamefont{Dijkstra}},
  \bibinfo{journal}{Phys. Rev. Lett.} \textbf{\bibinfo{volume}{98}},
  \bibinfo{pages}{095701} (\bibinfo{year}{2007}).

\bibitem[{\citenamefont{Berardi et~al.}(2003)\citenamefont{Berardi, Costantini,
  and Muccioli}}]{Berardi:2003}
\bibinfo{author}{\bibfnamefont{R.}~\bibnamefont{Berardi}},
  \bibinfo{author}{\bibfnamefont{A.}~\bibnamefont{Costantini}},
  \bibnamefont{and} \bibinfo{author}{\bibfnamefont{L.}~\bibnamefont{Muccioli}},
  \bibinfo{journal}{J. Chem. Phys.} \textbf{\bibinfo{volume}{126}},
  \bibinfo{pages}{044905} (\bibinfo{year}{2003}).

\bibitem[{\citenamefont{Bates}(2003)}]{Bates:2003}
\bibinfo{author}{\bibfnamefont{M.~A.} \bibnamefont{Bates}},
  \bibinfo{journal}{Chem. Phys. Lett.} \textbf{\bibinfo{volume}{368}},
  \bibinfo{pages}{87} (\bibinfo{year}{2003}).

\bibitem[{\citenamefont{Trukhina and Schilling}(2008)}]{trukhina:2008}
\bibinfo{author}{\bibfnamefont{Y.}~\bibnamefont{Trukhina}} \bibnamefont{and}
  \bibinfo{author}{\bibfnamefont{T.}~\bibnamefont{Schilling}},
  \bibinfo{journal}{Phys. Rev. E} \textbf{\bibinfo{volume}{77}},
  \bibinfo{pages}{011701} (\bibinfo{year}{2008}).

\bibitem[{\citenamefont{Oosawa and Asakura}(1954)}]{asakura.oosawa:1954}
\bibinfo{author}{\bibfnamefont{F.}~\bibnamefont{Oosawa}} \bibnamefont{and}
  \bibinfo{author}{\bibfnamefont{S.}~\bibnamefont{Asakura}},
  \bibinfo{journal}{J. Chem. Phys.} \textbf{\bibinfo{volume}{22}},
  \bibinfo{pages}{1255} (\bibinfo{year}{1954}).

\bibitem[{\citenamefont{Vrij}(1976)}]{vrij:1976}
\bibinfo{author}{\bibfnamefont{A.}~\bibnamefont{Vrij}}, \bibinfo{journal}{Pure
  Appl. Chem.} \textbf{\bibinfo{volume}{48}}, \bibinfo{pages}{471}
  (\bibinfo{year}{1976}).

\bibitem[{\citenamefont{Jungblut et~al.}(2007)\citenamefont{Jungblut, Tuinier,
  Binder, and Schilling}}]{Jungblut:2007}
\bibinfo{author}{\bibfnamefont{S.}~\bibnamefont{Jungblut}},
  \bibinfo{author}{\bibfnamefont{R.}~\bibnamefont{Tuinier}},
  \bibinfo{author}{\bibfnamefont{K.}~\bibnamefont{Binder}}, \bibnamefont{and}
  \bibinfo{author}{\bibfnamefont{T.}~\bibnamefont{Schilling}},
  \bibinfo{journal}{J. Chem. Phys.} \textbf{\bibinfo{volume}{127}},
  \bibinfo{pages}{244909} (\bibinfo{year}{2007}).

\bibitem[{\citenamefont{Gr{\"u}nwald et~al.}(2007)\citenamefont{Gr{\"u}nwald,
  Dellago, and Geissler}}]{gruenwald:2007}
\bibinfo{author}{\bibfnamefont{M.}~\bibnamefont{Gr{\"u}nwald}},
  \bibinfo{author}{\bibfnamefont{C.}~\bibnamefont{Dellago}}, \bibnamefont{and}
  \bibinfo{author}{\bibfnamefont{P.~L.} \bibnamefont{Geissler}},
  \bibinfo{journal}{J. Chem. Phys.} \textbf{\bibinfo{volume}{127}},
  \bibinfo{pages}{154718} (\bibinfo{year}{2007}).

\bibitem[{\citenamefont{Dijkstra et~al.}(2001)\citenamefont{Dijkstra, van Roij,
  and Evans}}]{dijkstra.roij.ea:2001}
\bibinfo{author}{\bibfnamefont{M.}~\bibnamefont{Dijkstra}},
  \bibinfo{author}{\bibfnamefont{R.}~\bibnamefont{van Roij}}, \bibnamefont{and}
  \bibinfo{author}{\bibfnamefont{R.}~\bibnamefont{Evans}},
  \bibinfo{journal}{Phys. Rev. E} \textbf{\bibinfo{volume}{63}},
  \bibinfo{pages}{051703} (\bibinfo{year}{2001}).

\bibitem[{\citenamefont{Low}(2002)}]{low:2002}
\bibinfo{author}{\bibfnamefont{R.~J.} \bibnamefont{Low}},
  \bibinfo{journal}{Eur. J. Phys.} \textbf{\bibinfo{volume}{23}},
  \bibinfo{pages}{111} (\bibinfo{year}{2002}).

\bibitem[{\citenamefont{Cuetos et~al.}(2007)\citenamefont{Cuetos,
  Martinez-Haya, and Lago}}]{Cuetos:2007}
\bibinfo{author}{\bibfnamefont{A.}~\bibnamefont{Cuetos}},
  \bibinfo{author}{\bibfnamefont{B.}~\bibnamefont{Martinez-Haya}},
  \bibnamefont{and} \bibinfo{author}{\bibfnamefont{S.}~\bibnamefont{Lago}},
  \bibinfo{journal}{Phys. Rev. E} \textbf{\bibinfo{volume}{75}},
  \bibinfo{pages}{061701} (\bibinfo{year}{2007}).

\bibitem[{\citenamefont{van~der Schoot}(1999)}]{schoot:1999}
\bibinfo{author}{\bibfnamefont{P.}~\bibnamefont{van~der Schoot}},
  \bibinfo{journal}{J. Phys. Chem. B} \textbf{\bibinfo{volume}{103}},
  \bibinfo{pages}{8804} (\bibinfo{year}{1999}).

\bibitem[{\citenamefont{Cuetos et~al.}(2008)\citenamefont{Cuetos, van Roij, and
  Dijkstra}}]{cuetos:2008}
\bibinfo{author}{\bibfnamefont{A.}~\bibnamefont{Cuetos}},
  \bibinfo{author}{\bibfnamefont{R.}~\bibnamefont{van Roij}}, \bibnamefont{and}
  \bibinfo{author}{\bibfnamefont{M.}~\bibnamefont{Dijkstra}},
  \bibinfo{journal}{Soft Matter} \textbf{\bibinfo{volume}{4}},
  \bibinfo{pages}{757} (\bibinfo{year}{2008}).

\bibitem[{\citenamefont{Nehring and Saupe}(1971)}]{Nehring:1971}
\bibinfo{author}{\bibfnamefont{J.}~\bibnamefont{Nehring}} \bibnamefont{and}
  \bibinfo{author}{\bibfnamefont{A.}~\bibnamefont{Saupe}}, \bibinfo{journal}{J.
  Chem. Phys.} \textbf{\bibinfo{volume}{54}}, \bibinfo{pages}{377}
  (\bibinfo{year}{1971}).

\bibitem[{\citenamefont{Lee and Meyer}(1986)}]{Lee:1986}
\bibinfo{author}{\bibfnamefont{S.-D.} \bibnamefont{Lee}} \bibnamefont{and}
  \bibinfo{author}{\bibfnamefont{R.~B.} \bibnamefont{Meyer}},
  \bibinfo{journal}{J. Chem. Phys.} \textbf{\bibinfo{volume}{84}},
  \bibinfo{pages}{3443} (\bibinfo{year}{1986}).

\bibitem[{\citenamefont{Vroege and Odijk}(1987)}]{Vroege:1987}
\bibinfo{author}{\bibfnamefont{G.~J.} \bibnamefont{Vroege}} \bibnamefont{and}
  \bibinfo{author}{\bibfnamefont{T.}~\bibnamefont{Odijk}}, \bibinfo{journal}{J.
  Chem. Phys.} \textbf{\bibinfo{volume}{87}}, \bibinfo{pages}{4223}
  (\bibinfo{year}{1987}).

\bibitem[{\citenamefont{Tjipto-Margo et~al.}(1992)\citenamefont{Tjipto-Margo,
  Evans, Allen, and Frenkel}}]{Tjipto:1992}
\bibinfo{author}{\bibfnamefont{B.}~\bibnamefont{Tjipto-Margo}},
  \bibinfo{author}{\bibfnamefont{G.~T.} \bibnamefont{Evans}},
  \bibinfo{author}{\bibfnamefont{M.~P.} \bibnamefont{Allen}}, \bibnamefont{and}
  \bibinfo{author}{\bibfnamefont{D.}~\bibnamefont{Frenkel}},
  \bibinfo{journal}{J. Phys. Chem.} \textbf{\bibinfo{volume}{96}},
  \bibinfo{pages}{3942} (\bibinfo{year}{1992}).

\bibitem[{\citenamefont{Lee}(1987)}]{Lee:1987}
\bibinfo{author}{\bibfnamefont{S.}~\bibnamefont{Lee}}, \bibinfo{journal}{J.
  Chem. Phys.} \textbf{\bibinfo{volume}{87}}, \bibinfo{pages}{4972}
  (\bibinfo{year}{1987}).

\end{thebibliography}
\end{document}